\documentclass[floatfix,aps,showpacs,twocolumn]{revtex4-1}

\usepackage{amsmath}
\usepackage{graphicx}
\usepackage{subfigure}

\newcommand{\ket}[1]{\mbox{$|#1\rangle$}}
\newcommand{\nv}{NV}
\newcommand{\re}{re}
\newcommand{\gsim}{\mathrel{\hbox{\rlap{\lower.5ex\hbox{$\sim$}} \kern-.3em \raise.4ex \hbox{$>$}}}}

\begin{document}

\title{Swarm optimization for adaptive phase measurements with low visibility}

\author{Alexander~J.~F.~Hayes}
\email{alexander.hayes@uqconnect.edu.au}
\affiliation{Department of Physics and Astronomy, Macquarie University, Sydney, NSW 2109, Australia.}
\author{Dominic~W.~Berry}
\affiliation{Department of Physics and Astronomy, Macquarie University, Sydney, NSW 2109, Australia.}

\begin{abstract}
Adaptive feedback normally provides the greatest accuracy for optical phase measurements.
New advances in nitrogen vacancy cent\re\ technology have enabled magnetometry via individual spin measurements,
which are similar to optical phase measurements but with low visibility.
The adaptive measurements that previously worked well with high-visibility optical interferometry break
down and give poor results for nitrogen vacancy cent\re\  measurements. We use advanced search techniques based
on swarm optimisation to design better adaptive measurements that can provide improved measurement
accuracy with low-visibility interferometry, with applications in nitrogen vacancy cent\re\  magnetometry.
\end{abstract}


\maketitle

\section{Introduction}
\label{intro}
An important feature of quantum mechanics is that it imposes fundamental limits on how accurately physical quantities can be measured.
On the other hand, by taking advantage of the features of quantum mechanics, one can perform measurements that are far more accurate than would otherwise be possible.
Many types of precision measurement use a form of interferometry.
One particular example is optical interferometry, which is used to measure distance via changes in phase.
In this case, if one uses $N$ independent photons, then the accuracy of the phase measurement scales as $1/\sqrt{N}$.
On the other hand, if one takes advantage of quantum mechanics by using $N$ photons in a special entangled state, then the accuracy can scale as $1/N$, yielding vastly higher accuracy for large $N$ \cite{caves,yurke,giov04}.

A commonly considered entangled state is the NOON state, where $N$ photons are in a superposition of being all in one arm of the interferometer or the other \cite{sanders,bollinger,lee}.
These states have the advantage that they provide accuracy scaling as $1/N$, with the drawback that the phase needs to be initially known to this accuracy; otherwise the measurement is ambiguous.
One way of interpreting the accuracy is that $m$ measurements with $m$ copies of this state will yield accuracy $1/(N\sqrt m)$.
An alternative approach is to combine measurements using NOON states with many different values of $N$.
The measurements with smaller values of $N$ are used to resolve the ambiguity in the measurements with larger values of $N$.
This is the approach used in Refs.~\cite{xiang11,higg09}.
Instead of considering NOON states, one can instead consider multiple passes through a phase shift~\cite{higg07}.

Another area of interferometry is that using transitions in atomic or solid-state systems.
This can be used for time standards or frequency measurement, or probing physical quantities that affect the frequency.
The particular application we consider here is the use of a single nitrogen-vacancy (NV) cent\re\ in diamond to probe the magnetic field at the nanoscale \cite{taylor08,maze08,bala08,degen08,stein10}.
Experiments in such systems have demonstrated sensitivity of $\sim 3$ nT with a spatial resolution of $\sim 5$ nm.
The \nv\ cent\re\ has excellent properties for magnetic sensing, because it can be individually addressed and maintains spin coherence for a significant period of time at room temperature.

A fundamental difficulty with these measurement techniques is that the measurement signal has periodic modulation.
The measurement time is either restricted to half an oscillation period, or the magnetic field range must be known accurately in advance.
This is essentially the same problem as occurs when performing phase measurements with NOON states, and it is therefore natural to apply the same measurement schemes as were developed for NOON states.
This is what was proposed in Ref.~\cite{said11}, and it was experimentally demonstrated in Refs.~\cite{wald12,nusran,haberle}.
For this system the interpretation is that the measurements provide a high dynamic range, rather than a quantum improvement as in the case of NOON states.

In optical interferometry, it has been found that adaptive measurements are typically the most accurate \cite{berry00,berry01,berry01b,higg07,xiang11,yone12,spang}.
That is, information from early parts of the measurement are used to adjust how the measurement is performed.
In fact, for measurements on a single mode, nonadaptive measurements are unable to achieve better than $1/\sqrt{N}$ scaling even for highly nonclassical states \cite{wiseman}.
In contrast, for measurements on multiple time modes, such as measurements on multiple NOON states, nonadaptive measurements can yield the same accuracy scaling as adaptive measurements \cite{higg09,berry09}.

An important difference between optical measurements and measurements with \nv\ cent\re s is that the \nv\ cent\re\ measurements have lower visibility.
There is both a reduced initial visibility, and the visibility decreases exponentially with interaction time due to decoherence.
In the experiment in Ref.~\cite{wald12} the initial visibility is only about 80\%, whereas visibilities in optical interferometry are often 98\% or better \cite{higg07}.
It turns out that when the visibility is this low, the adaptive measurement schemes that have been developed in previous work become very inaccurate \cite{said11}.
This is why nonadaptive measurements were used in Refs.~\cite{wald12,nusran}.
Even the nonadaptive measurements have somewhat poor performance for low visibility.
It should be possible for some adaptive method to provide improved performance, because adaptive measurements are more general than nonadaptive measurements.

The adaptive techniques that were considered for \nv\ cent\re\ magnetometry before were those based on the technique from Refs.~\cite{berry00,berry01}.
This technique only optimises the measurement locally, in the sense that it minimises the variance after the next detection.
An alternative technique is to use global optimisation to minimise the final variance at the end of the measurement.
Such an optimisation is far more challenging, but particle swarm optimisation techniques have been found to be effective for single-time-mode measurements \cite{hent10,hent11}.
An alternative technique for measurements with multiple NOON states of different sizes was proposed by Cappellaro \cite{cappe12}.
Here we combine particle swarm optimisation techniques with the method of Cappellaro to find adaptive measurements that provide improved performance for \nv\ cent\re\  magnetometry.

\section{Ramsey interferometry}
\label{sec:ram}
In this section we summarise Ramsey interferometry, and how it is used with \nv\ cent\re s.
The probe state is prepared in the superposition state $(\ket{0} + \ket{1})/\sqrt 2$.
Given that the energy level splitting is $\Delta E$, the state evolves over time $t$ to $(\ket{0} + e^{-it\Delta E/\hbar}\ket{1})/\sqrt 2$.
In the case we are interested in, the energy level splitting is due to the different spin states, and is proportional to the magnetic field.
Therefore the state is $(\ket{0} + e^{-2it\gamma B}\ket{1})/\sqrt 2$, where $\gamma$ is the gyromagnetic ratio.
For readout, the state is measured in the basis $(\ket{0} \pm \ket{1})/\sqrt 2$.
The probabilities of the measurement results are then
\begin{equation}
P(\pm|B) = \frac 12 \left[ 1\pm \cos(2t\gamma B)\right].
\end{equation}

\begin{figure}
\begin{center}
\includegraphics[width=7.5cm]{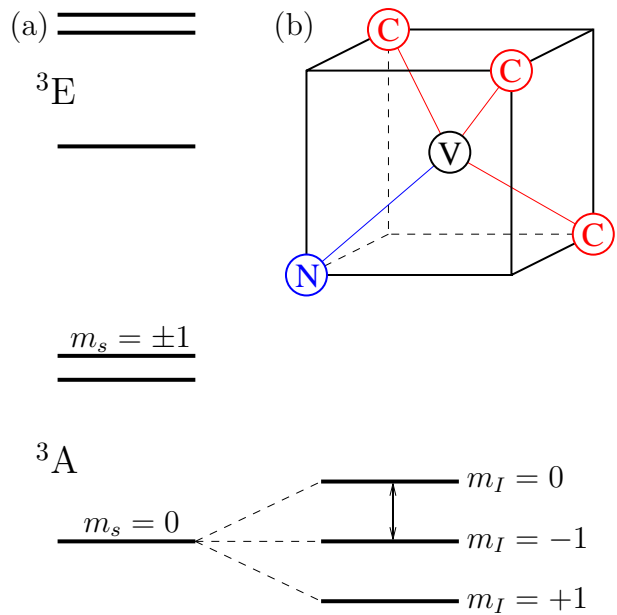}
\put(-230,216){\large (a)}
\put(-130,216){\large (b)}
\put(-220,190){\Large $^3$E}
\put(-220,50){\Large $^3$A}
\put(-207,98){\large $m_s=\pm 1$}
\put(-202,28){\large $m_s=0$}
\put(-57,45){\large $m_I=0$}
\put(-57,22){\large $m_I=-1$}
\put(-57,-1){\large $m_I=+1$}
\caption{The energy level structure of an \nv\ cent\re.
(a) There are two sets of energy levels, $^3$A and $^3$E.
We consider the electronic spin state $m_s=0$ of $^3$A, which has the further hyperfine splitting into $m_I=0$ and $\pm 1$.
(b) The structure of the \nv\ cent\re, consisting of a substitutional nitrogen impurity N (blue circle) adjacent to a vacancy V (black circle), tetrahedrally arranged with respect to each other and to the nearby carbon atoms C (red circles).}
\label{levels}
\end{center}
\end{figure}

The negatively charged \nv\ cent\re, denoted NV$^-$, has an energy level diagram as shown in Fig.~\ref{levels}.
There are two sets of energy levels, $^3$A and $^3$E, labelled by the irreducible representations of the symmetry group of the defect cent\re.
There are $3$ allowable electronic spin states for each, with $m_s=0$, and $\pm 1$.
The electronic spin states $m_s=\pm 1$ for $^3$A were used in Ref.~\cite{nusran}.
In the case of nitrogen-14, the energy level $m_s=0$ for $^3$A is further split into nuclear energy levels $m_I=0$, and $\pm 1$.
The nuclear spin states $m_I=0$ and $-1$ were used in Ref.~\cite{wald12}.
These were then mapped to electronic spin states for readout.

In order to obtain the superposition state, Ref.~\cite{wald12} initialised the system in the $m_I=0$ state,
then used a $\pi/2$ pulse to obtain a superposition of $m_I=0$ and $-1$.
After allowing the system to accumulate a phase depending on the magnetic field, another $\pi/2$ pulse was used, then
the system was measured in the $m_I=0$ and $-1$ basis.
The advanced phase estimation algorithms developed in optics depend on using a controllable phase, that
is either controlled in a predetermined way (in the nonadaptive case), or based on the results of measurements (in the adaptive case).
In optics this controlled phase can be implemented by putting a phase modulator in one arm of an interferometer.
In the Ramsey interferometry experiment in Ref.~\cite{wald12} this controlled phase was implemented by changing the phase of the second $\pi/2$ pulse.
By changing the phase of the pulse, the system could be rotated around the $x$ or $y$ axis of the Bloch sphere.

In general, consider a rotation of the form
\begin{equation}
R(\theta)=\exp[i(\sigma_x \sin\theta + \sigma_y \cos\theta)\pi/4].
\end{equation}
Performing this rotation followed by a measurement yields the probabilities
\begin{equation}
P(\pm|B) = \frac 12 \left[ 1\pm \cos(2t\gamma B-\theta)\right].
\end{equation}
Hence this technique provides a controlled phase that is mathematically equivalent to that used in optics.

For the phase measurement technique of Refs.~\cite{said11,wald12,nusran}, the times used are multiples of a base time, $\tau$.
It is therefore convenient to define a phase
\begin{equation}
\phi := 2 \gamma B \tau .
\end{equation}
The phase shifts will be of the form $2^k\phi$.
No multiples of $\phi$ smaller than $1$ will be considered, so $\phi$ is measured in the range $(-\pi,\pi]$.
The corresponding range for measurement of $B$ is $(-B_{\rm max},B_{\rm max}]$, where
\begin{equation}
B_{\rm max} := \frac {\pi}{2\gamma\tau}.
\end{equation}

An important feature of Ramsey interferometry is the visibility of the interference.
It starts significantly less than $100\%$, then exponentially decays with time due to decoherence.
Taking this into account, the probability distribution for the measurement results is \cite{said11}
\begin{equation}
\label{eq:probdist}
P(\pm | \phi ) = \frac{1}{2}[1 \pm f_d e^{-2^k \tau / T_2}\cos\left(2^k \phi - \theta \right) ],
\end{equation}
where $f_d$ is a factor representing the visibility of the measurement, and $T_2$ is the transverse spin coherence time.
The time $T_2$ indicates the decay rate of spin coherence in the system, and sets an effective limit on the maximum duration of measurements on the system.

If one were to just use all measurements with the same interaction time, $\tau$, then for a \emph{total} interaction time $T$ the phase variance would be no less than
$1/N$, where $N:=T/\tau$.
This corresponds to an uncertainty in $B$ lower bounded as
\begin{equation}
\Delta B \ge \frac 1{2\gamma\tau\sqrt{N}}.
\end{equation}
We can reduce the variance by increasing $\tau$, but that also reduces the range of the measurement.
The dynamic range (the ratio of the uncertainty to the maximum magnetic field that can be detected) is lower bounded as
\begin{equation}
\label{eq:lim1}
\frac{\Delta B}{B_{\rm max}} \ge \frac{\pi}{\sqrt{N}}.
\end{equation}

In contrast, when we use multiple interaction times, the phase variance is lower bounded as approximately $(\pi/N)^2$ (the exact lower bound is in Eq.~\eqref{eq:holbnd}).
The corresponding lower bound to the dynamic range is then
\begin{equation}
\label{eq:lim2}
\frac{\Delta B}{B_{\rm max}} \gsim \frac 1N.
\end{equation}
That is, we potentially obtain a square improvement in the dynamic range by using measurement techniques with multiple interaction times.
This is analogous to the improvement from the standard quantum limit to the Heisenberg limit with NOON states.

\section{Measurement techniques and analysis}
\label{meth}

\subsection{Bayesian estimation}

In order to determine estimates of the phase (and thereby the magnetic field),
it is convenient to use Bayesian estimation to calculate the probability distribution for the system phase based on successive measurement results \cite{berry00,berry01}.
This enables one to calculate the exact phase variance for a given measurement scheme \cite{berry00,berry01}.
It is assumed that the magnetic field is initially unknown, other than being confined to the range $(-B_{\rm max},B_{\rm max}]$.
This means that there is no initial knowledge of the phase, so the prior distribution is
\begin{equation}
P(\phi | \vec u_0) = \frac{1}{2 \pi}.
\end{equation}
Here the notation that is used is that the successive measurement results are $u_1$, $u_2$, and so forth, and a vector of $n$ measurement results is $\vec u_n:=(u_1,\ldots,u_n)$.
The initial vector of zero length before any measurement results are taken is $\vec u_0$.

Each interferometric measurement provides additional information about the phase $\phi$, and therefore the magnetic field.
This information is quantified by using Bayes' rule to update the probability distribution as
\begin{equation}
P(\phi | \vec u_n) \propto P(u_n | \phi) P(\phi | \vec u_{n-1}),
\label{bayesequ}
\end{equation}
where $u_n$ is the outcome of the most recent measurement.
Note that the denominator of the Bayesian function has not been
explicitly given here as it is independent of the measurement outcome, and serves only as a normalising factor for the probability distribution.
The formula to use for $P(u_n | \phi)$ is given in Eq.~\eqref{eq:probdist}.

\subsection{The Fourier Series Representation}

As the Bayesian probability distribution is a product of sinusoids, it is convenient to use the Fourier series of the probability distribution function to represent it.
We write the general form of the probability distribution as
\begin{equation}
P(\phi) = \sum_{w=-\infty}^{\infty} b_w e^{i w \phi}.
\end{equation}
Note that, in practice, the sum need not be taken to infinity, because the probability distribution is a product of a limited number of sinusoids.

Combining the representation above with Eq.~\eqref{eq:probdist}, it is possible to derive the simple update rule for the coefficients of the Fourier series
\begin{equation}
b_{w}^{n} = \frac{1}{2}b_{w}^{n-1}+\frac{u_nV}{4}b_{w-2^k}^{n-1}e^{-i \theta}+\frac{u_nV}{4}b_{w+2^k}^{n-1}e^{i \theta},
\end{equation}
where $V:= f_d e^{-2^k \tau / T_2}$.
Note that this formula is only applicable when the system is restricted to integer multiples of the interaction time, which will be the case for the metrology methods considered within this paper.
Numeric simulations of the phase estimation process may be performed by tracking the coefficients of the probability distribution and updating them for each measurement using the formula given above.

To estimate the accuracy of the phase measurement, it is convenient to use the Holevo variance \cite{holevo}
\begin{equation}
V_H := \frac 1{|\langle e^{i(\hat\phi-\phi)}\rangle|^2} -1,
\end{equation}
where $\hat\phi$ is the estimate of the phase.
To account for biased estimates of the phase one can use $\langle \cos(\hat\phi-\phi)\rangle$ in place of $\langle e^{i(\hat\phi-\phi)}\rangle$.
The Holevo variance is close to the usual variance for narrowly peaked distributions.
For phase an advantage of the Holevo variance is that it is naturally modulo $2\pi$.
That is, a phase close to $-\pi$ is regarded as also close to $\pi$.
In the case where the phase is used for measuring the magnetic field, this is a little problematic, because an estimate near $-B_{\rm max}$ would be regarded as accurate if the actual magnetic field is near $B_{\rm max}$.

To avoid the problem of having a large error in the estimate of the field near $\pm B_{\rm max}$, one can consider an initial probability distribution for the magnetic field that is in the range $(-B'_{\rm max},B'_{\rm max}]$, where $B'_{\rm max}=B_{\rm max}-c\Delta B$.
Here $c$ is a constant chosen such that the probability of the error (in the magnetic field estimate) being larger than $c\Delta B$ is negligible.
Then the actual dynamic range would be $\Delta B/B'_{\rm max}$, but for large dynamic range this will be close to $\Delta B/B_{\rm max}$.
We do not consider that correction in this work, because we consider large dynamic range where it is negligible.

Another feature of the Holevo variance is that it means only one coefficient of the Fourier series for the probability distribution is important.
First, the optimal phase estimate to use is $\hat\phi=\arg(b_{-1})$ \cite{berry01}.
Second, the Holevo variance can be determined exactly by summing the value of $|b_{-1}|$ over all combinations of measurement results.
See Ref.~\cite{berry09} for explanation of how this is done.
It is therefore unnecessary to track all nonzero coefficients, and we can restrict ourselves to recording those that contribute to the final value of $b_{-1}$.
This considerably simplifies the calculation.
The calculation can be further simplified by noting that the probability distribution is real, so $b_{-w}=b_w^*$, and only half the coefficients need be stored.

The lower bound to the Holevo variance is \cite{berry00,luis}
\begin{equation}
\label{eq:holbnd}
V_H \ge \tan^2\left[ \frac{\pi}{N+2} \right].
\end{equation}
This bound is slightly less than $(\pi/N)^2$, but it approaches $(\pi/N)^2$ in the limit of large $N$.
The lower bound $1/N$ given in Sec.~\ref{sec:ram} for measurements with all the same interaction time was also in terms of $V_H$.
In that case it is not a tight lower bound, and there does not appear to be a known analytic expression for the tight bound.

\subsection{Phase Measurement Protocols}

There are several different aspects of the phase estimation procedure that can be controlled with the aim of achieving the best possible estimate.
One can obtain improved performance with entangled states;
for example, the equivalent of optical NOON states would be $N$ entangled \nv\ cent\re s.
Entangled \nv\ cent\re s have been demonstrated \cite{dolde}, but are not yet at the stage where they can effectively be used for magnetometry.
For this reason we consider the improvement of the choice of interaction time and control phase $\theta$.

The procedure to perform the phase is to begin with the longest viable interaction time, and then systematically reduce it by a factor of $2$, performing multiple repetitions for each interaction time.
In the following we use the terminology ``detection'' for the individual measurements, to distinguish them from the overall measurement based on the combination of these individual detections.
The initial interaction time is $2^K\tau$, and the sequence of multipliers is $2^K, 2^{K-1},\ldots, 2^1, 2^0$.

The number of detections used for the longest interaction time is denoted $G$, and each time the interaction time is halved the number of detections is increased by $F$.
Therefore, for interaction time $2^k\tau$, the number of detections is
\begin{equation}
M_k = G+F(K-k).
\end{equation}
The total interaction time used is then \cite{said11}
\begin{equation}
T=\tau [G(2^{K+1}-1)+F(2^{K+1}-2-K)].
\end{equation}
This sequence is similar to that in Refs.~\cite{said11,wald12,nusran}, and to the optical technique in Ref.~\cite{higg07}.

Although it was found to be possible to take $F=0$ for adaptive measurements~\cite{higg07}, for nonadaptive measurements it is necessary to take $F>0$ \cite{higg09}.
Reference \cite{said11} found that the nonadaptive measurements gave better results than the adaptive technique of Ref.~\cite{higg07} for poor visibility.
Nevertheless, if the initial visibility was too low, the nonadaptive measurements were still quite poor.
The experiments in Refs.~\cite{wald12,nusran} addressed this by increasing the number of detections for each interaction time.
Reference~\cite{wald12} used $G=36$ and $F=8$, whereas Ref.~\cite{nusran} used $G=F=9$.

\begin{figure}
\begin{center}
\includegraphics[width=7cm]{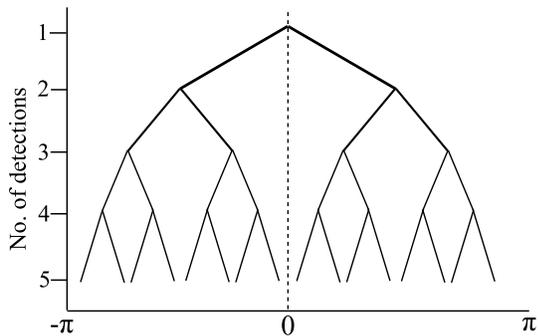}
\caption{An illustration of a basic symmetric protocol decision tree.
The horizontal axis indicates the phase offset from its initial setting.}
\label{ptree}
\end{center}
\end{figure}

The key difference in our approach is how we adjust the control phase $\theta$ based on the measurement results.
In general, the strategy for choosing the control phase based on measurement results can be described by a binary decision tree, as in Fig.~\ref{ptree}.
The approach used in Ref.~\cite{hent10} was to choose the size of the step dependent on the detection result and the number of previous detections.

An alternative approach is to update the controlled phase based on the estimate of the system phase.
This approach is essentially an adaptive homodyne measurement, similar to Refs.~\cite{wheat10,yone12}.
Here, after detections with interaction times down to $2^k\tau$, the phase information has only been obtained modulo $2\pi/2^k$.
This means that the only nonzero Fourier coefficients $c_w$ are those where $w \mod 2^k=0$.
This is advantageous for computation, because it means that only those coefficients need be recorded.
However, it means that one cannot use $c_{-1}$ for an estimate of the phase (unless $k=0$).
Instead one can use $c_{-2^k}$ for an estimate of the phase modulo $2\pi/2^k$; that is, it is an estimate of $2^k\phi$.
Taking $\theta$ to be an estimate of $2^k\phi$ plus $\pi/2$ yields the point where the probabilities in Eq.~\eqref{eq:probdist} are most sensitive to $\phi$.

Nevertheless, this adaptive homodyne approach performs very poorly with reduced visibility of the interference.
Recently, a variation was proposed by Cappellaro \cite{cappe12} for this type of phase estimation, in which the controlled phase is updated only after each change of interaction time.
That is, when the interaction time is changed from $2^{k}\tau$ to $2^{k-1}\tau$, the controlled phase is taken to be
\begin{equation}
\label{eq:capp}
\theta = \frac{1}{2}\arg \left(b_{-2^k}\right).
\end{equation}
An estimate of $2^k\phi$ is obtained as $\arg(b_{-2^k})$.
This is then divided by 2 to obtain an estimate of $2^{k-1}\phi$, as is needed because one has $2^{k-1}\phi-\theta$ in the argument of the cosine in Eq.~\eqref{eq:probdist}.

Note that there is ambiguity modulo $\pi$ in the estimate of $2^{k-1}\phi$, but this is unimportant because adding $\pi$ yields equivalent results.
Specifically, adding $\pi$ exchanges the probabilities for detection results $u=+1$ and $u=-1$.
Hence, if $\pi$ is added and the detection result is $u=-1$, the following calculations are the same as if $\pi$ were not added and the detection result was $u=+1$.
This means that the same variance must be obtained regardless of whether $\pi$ is added.

Note that the formula \eqref{eq:capp} does not have $\pi/2$ in it, so it does not give the point where the probabilities are most sensitive to $\phi$.
However, it turns out that when $\pi/2$ is not used, this is the phase that minimises the variance after the next detection (after the change in interaction time).
This can be verified using the formula in Ref.~\cite{berry01}.

\section{Optimisation Algorithms and Techniques}

\begin{figure}
\begin{center}
\includegraphics[width=8cm]{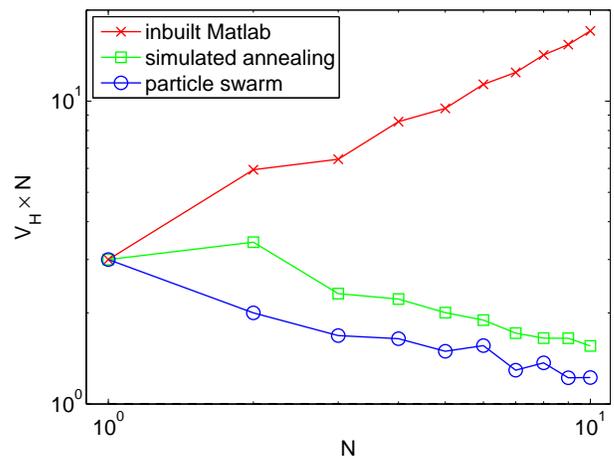}
\caption{A performance comparison of different optimisation methods for an example simulation.
The methods shown are Matlab's inbuilt non-linear programming suite, a simulated annealing approach, and the particle swarm algorithm.
Each was given the same runtime for a phase estimation simulation task.}
\label{bmar}
\end{center}
\end{figure}

To find an improved adaptive protocol for this type of phase estimation system, we applied numerical optimisation to minimise the variance estimated by repeated simulations.
Several different optimisation algorithms were tested, included non-linear programming methods and simulated annealing, however we found that the procedure which gave the best result in a reasonable amount of time was particle swarm optimisation (PSO), which was also used in \cite{hent10, hent11}.

A comparison of some example results from the three different techniques are shown in Fig.~\ref{bmar}.
These calculations were for all the same interaction time $\tau$.
Because $N$ is the total interaction time divided by $\tau$, here it is the number of detections.
For this example $T_2$ was taken to be infinite for simplicity.
Approximately the same calculation times were allowed in each case.
Matlab's inbuilt optimisation routine produced the largest variances.
Simulated annealing produced somewhat improved results, and PSO produced the smallest variance.

The PSO algorithm uses a group of interacting `particles' to search the phase space of the problem. Each particle has a velocity which is updated with semi-random adjustments that are dependent on best values found by both the individual particle and by the entire group. The formula for the velocity update is
\begin{equation}
\vec v' = \chi [\vec v + c_g r_g (\vec x_g - \vec x) + c_l r_l (\vec x_l - \vec x)],
\end{equation}
where $\vec x$ is the particle position, $\vec v$ is the particle velocity, and $\vec v'$ is the updated velocity.
The variables $\vec x_g$ and $\vec x_l$ are the currently known global and local optimums, respectively,
$\chi$ is an overall damping factor to ensure convergence, $c_g$ and $c_l$ are set weightings for the current global and local optimums respectively, and $r_g$ and $r_l$ are random numbers that are chosen uniformly in the interval $[0,1)$ at each step.
We used $\chi=0.729$ and $c_g = c_l = 2.05$.
These are the weighting values recommended by Kennedy and Eberhart \cite{keneber}, the developers of the algorithm, and we found these to perform relatively well.

The space consisted of the phase increments after each detection.
That is, after each detection there was an increment in the controlled phase $\theta$ that depended on the number of the stage and the detection result, but not on prior results.
The position is the set of these phase increments.
The initial positions were chosen uniformly at random in the space.
The initial velocities were chosen uniformly at random, with the maximum velocity corresponding to half the size of the space per iteration.
The maximum velocity tended to act as a soft bound on the area covered, as the particles would usually stay within an area of dimensions that were roughly three times the maximum velocity.

We used fixed boundaries, since we were examining a domain equivalent to one full phase rotation.
The phase space had reflective boundaries to prevent a trapping effect near the edges.
The simulations used 10 particles;
we tested higher numbers of particles, but it did not improve the performance. 
The maximum number of iterations was set at 300 to allow the particles to fully converge.

\section{Simulation and Optimisation of Measurement Protocols}
\label{results}

\begin{figure}
    \centering
    \subfigure    {\includegraphics[width=8cm]{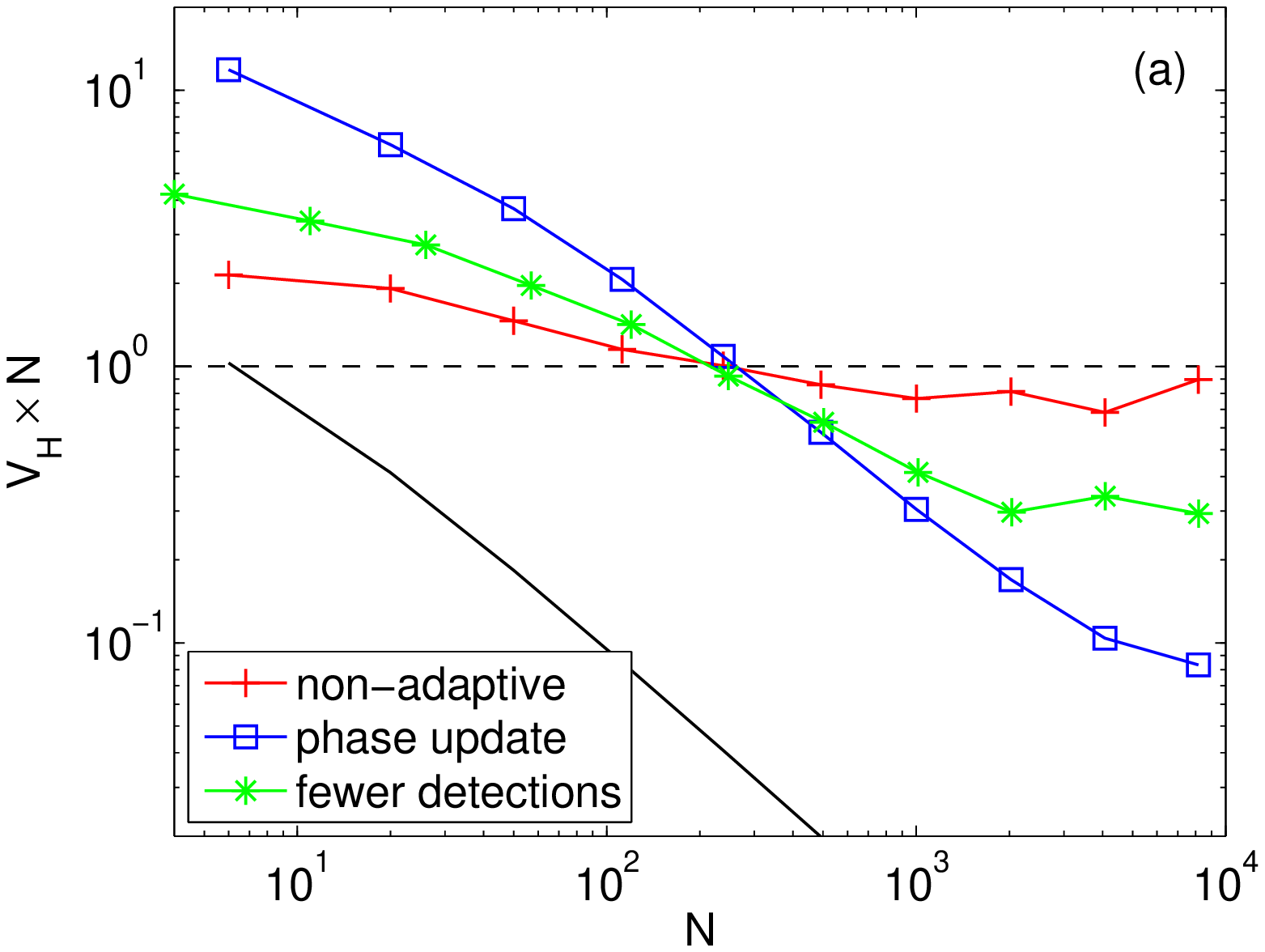}}    \\
    \subfigure    {\includegraphics[width=8cm]{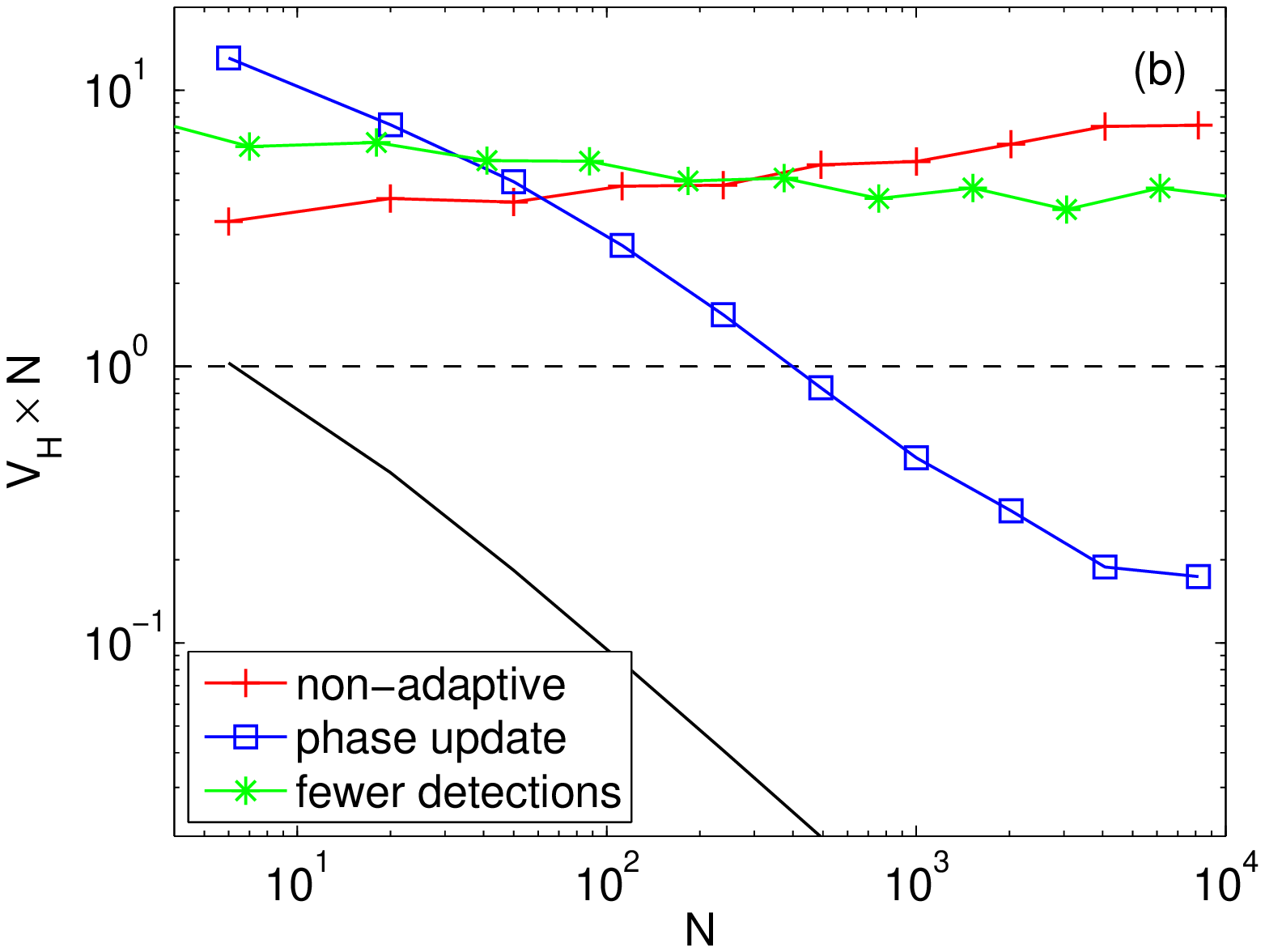}}
    \caption{A comparison of variances for different methods versus $N$ (the ratio of the total interaction time to $\tau$).
Shown are lines for a nonadaptive decision-tree protocol, a phase update protocol based on Cappellaro's work, and a phase update protocol with a reduced number of detections ($G=2$, $F=1$).
The limit for a single pass time is shown as the dotted line, and the limit for multiple pass times is shown as a solid line.
Parts (a) and (b) are with initial visibilities of $f_d=95\%$ and $f_d=85\%$, respectively.
All protocols shown other than the reduced detections phase update protocol use $G=6$ and $F=2$.}
    \label{fig:non}
\end{figure}

\begin{figure}
    \centering
    \subfigure  {\includegraphics[width=8cm]{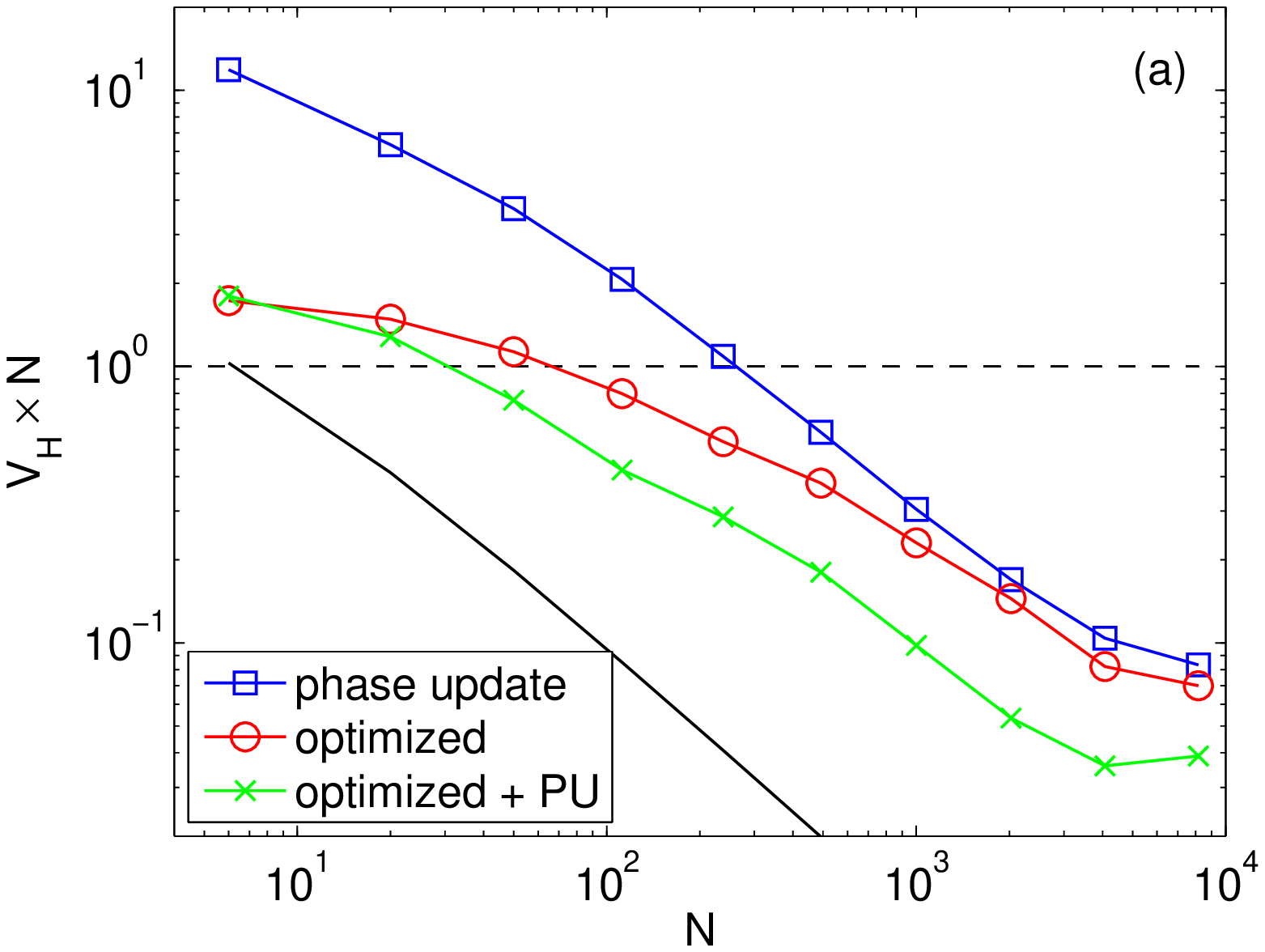}}    \\
    \subfigure  {\includegraphics[width=8cm]{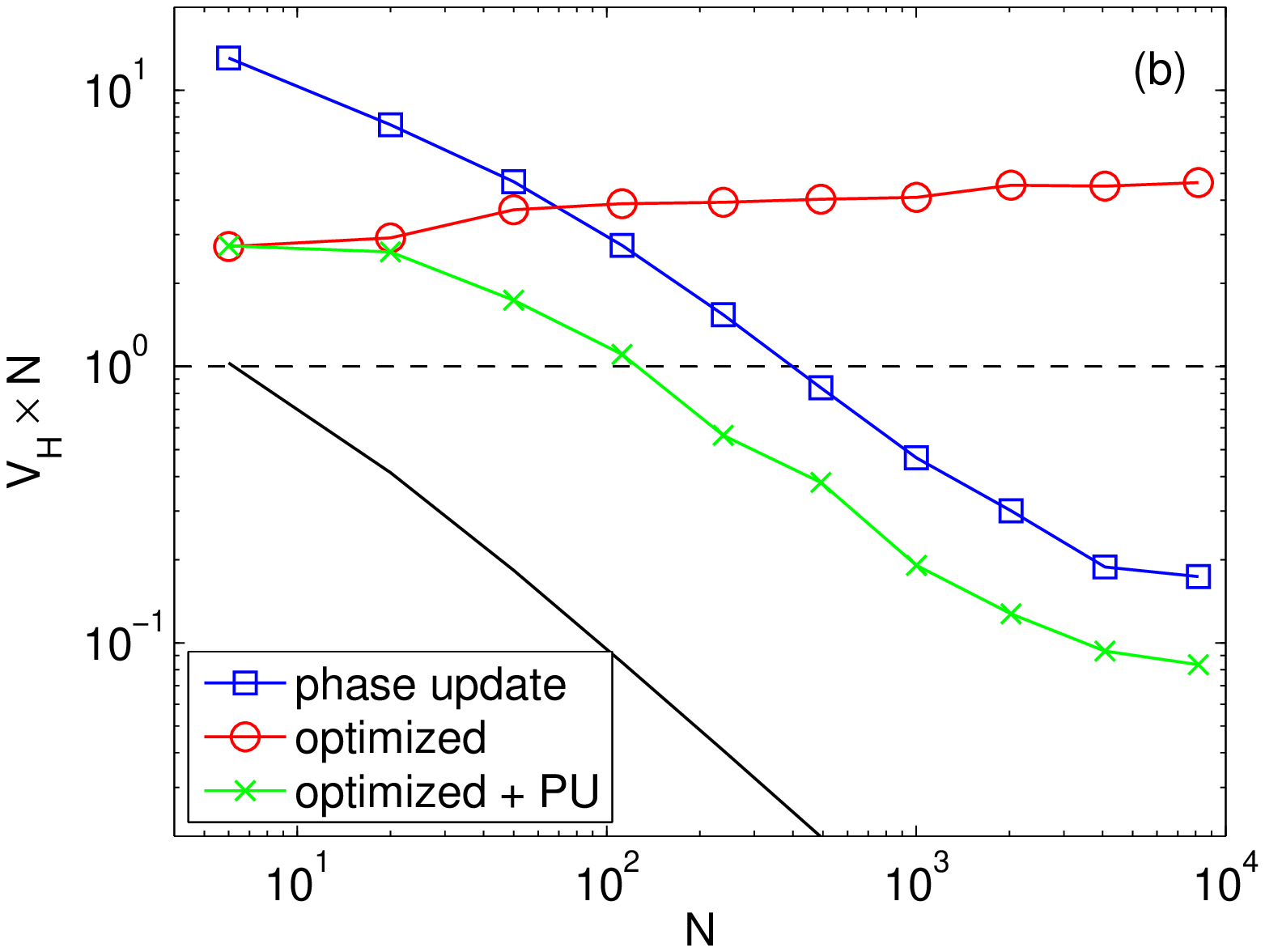}}
    \caption{A comparison of variances for different methods versus $N$ (the ratio of the total interaction time to $\tau$).
Shown are lines for a phase update protocol based on Cappellaro's work, an optimised decision-tree protocol and a hybrid protocol combining an optimised decision-tree with phase updates.
The limit for a single pass time is shown as the dotted line, and the limit for multiple pass times is shown as a solid line.
Parts (a) and (b) are with initial visibilities of $f_d=95\%$ and $f_d=85\%$, respectively.}
    \label{fig:opt}
\end{figure}

We ran simulations to compare the performance of several different protocols.
In each case, we used a sample size of at least $2^{16}$ ($65536$),
intended to be sufficiently large to ensure accurate results while still being tractable for the computation equipment we used.
In order to reduce the random variation just due to the sampling, we used the same set of random numbers each time.
At the end we used a different set of random numbers to obtain the final estimate of the variance.
This is to ensure that the small variance was not just an artefact of the particular random numbers used.

For our optimised adaptive stepping, the optimisation program allowed for asymmetric steps, meaning that the size and direction of a step after a detection outcome is independent of the step resulting from the alternative outcome.
This also means that we were required to optimise over twice as many variables.

A further consideration was the number of detections for each interaction time.
For most of the results, we used the values $G = 6$ and $F = 2$, which were found to be sufficiently large to give improved scaling at high visibility, while also being useful for comparison with results from other work, such as \cite{said11}.
Smaller detection numbers were found to perform poorly for all protocols.

The results are shown in Figs.~\ref{fig:non} and \ref{fig:opt}, with parts (a) and (b) corresponding to initial visibilities of $f_d=95\%$ and $f_d=85\%$, respectively.
In these figures we took $T_2=10^3 \tau$.
In Fig.~\ref{fig:non}, the results for nonadaptive measurements are shown, as was used in Ref.~\cite{said11}.
For initial visibility of $f_d=95\%$ the product $V_H N$ decreases with $N$, whereas for $f_d=85\%$ the product $V_H N$ slightly increases with $N$.
These results are similar to those in Ref.~\cite{said11} (see Fig.~3 of that work).

Note that $V_H$ is proportional to the square of the dynamic range; that is,
\begin{equation}
V_H = \pi^2 \left( \frac{\Delta B}{B_{\rm max}} \right)^2.
\end{equation}
This means that the limit to the dynamic range for equal measurement times \eqref{eq:lim1} corresponds to $V_H N$ being constant.
Similarly, the ultimate limit to the dynamic range in \eqref{eq:lim2} corresponds to $V_H N$ scaling as $1/N$.
These limits are shown in Figs.~\ref{fig:non} and \ref{fig:opt} for reference.
It can be seen that for the nonadaptive measurement the results are quite similar to what could be obtained for equal interaction times.

An improved result is obtained by only updating the controlled phase each time the interaction time is reduced, according to the Cappellaro approach discussed above.
This yields a result that beats the equal-interaction-time limit for visibility of both $95\%$ and $85\%$.
It has a scaling quite similar to the ultimate limit, though it is a significant distance above the line (about a factor of $10$).
The alternative values $G=2$, $F=1$, which yield fewer detections for each interaction time, also give much higher variances.

The phase variances obtained using the PSO method to choose the controlled phases are shown in Fig.~\ref{fig:opt}.
This improves on the phase update based on Cappellaro's technique for visibility of $95\%$, but not for $85\%$ visibility.
To improve the performance, we tested a combined protocol using both optimised adaptive steps and a phase update based on the Bayesian distribution at each change in interaction time.
This method outperformed all others for both visibilities examined.

Although this optimised protocol shows scaling similar to the fundamental limit for moderate values of $N$, it has an upturn for the largest values of $N$.
This is similar to the result found in Ref.~\cite{said11}, and is again due to the fixed coherence time $T_2$.
The largest value of $N$ shown is $8164$, which corresponds to the longest interaction time being $512\tau$.
In comparison, the coherence time was $T_2=10^3\tau$, so the upturn occurs where the longest interaction time is comparable to $T_2$.
This is similar to the result in Ref.~\cite{said11}.
Hence, although our technique can provide improved performance for reduced \emph{initial} visibility, it is still limited by the coherence time, which is a fundamental limitation.

\section{Conclusions}
\label{conc}
Adaptive measurements usually give more accurate results than nonadaptive measurements, but previous adaptive techniques gave poor results for measurements with reduced visibility.
The nonadaptive measurements gave better results, but when the initial visibility is too small even nonadaptive measurements perform poorly.
A partially adaptive scheme was proposed by Cappellaro, which provided some improvement \cite{cappe12}.

Here we have used PSO to find a hybrid technique, that uses the adaptive step from Cappellaro, as well as optimising the other adaptive steps in the measurement.
This technique substantially improves on the technique of Cappellaro, as well as the nonadaptive technique and the earlier adaptive technique from Ref.~\cite{higg07}.
As a result, this hybrid technique should give a substantially improved result when applied to \nv\ cent\re\ magnetometry.

It is also possible to use PSO to search for an adaptive technique, without also using the Cappellaro update step.
In principle this should give the best result, but in practice it did not converge sufficiently well.
For initial visibility of $95\%$, it still gave an improvement over the Cappellaro technique, but it performed very poorly for $85\%$ visibility.

For these results we have not optimised over the number of detections for each interaction time.
There is the possibility that modifying the number of detections for each interaction time may yield a result that is further improved.
Further work remains to be done to determine the best numbers of detections to use, and the dependence of this number on the visibility.

\section{Acknowledgements}
\label{ack}
We would like to acknowledge helpful discussions with Jason Twamley and Alexandr Sergeevich. DWB is funded by an ARC Future Fellowship (FT100100761).

\end{document}